\RequirePackage{fix-cm}
\documentclass[%
 aip,
 jmp,%
 amsmath,amssymb,
 reprint,%
]{revtex4-2}         
\usepackage{graphicx}
\usepackage[utf8]{inputenc}
\usepackage[T1]{fontenc}
\usepackage{tabularx}
\usepackage{colortbl}
\usepackage{xcolor}
\usepackage{multirow}
\usepackage{float}
\usepackage{amsmath}
\usepackage{amssymb}
\usepackage{hyperref}
\usepackage{lipsum} 

\begin{document}
\title{Diversity of neuronal activity is provided by hybrid synapses 
}


\author{Kesheng Xu}
 \affiliation{Faculty of Science, Jiangsu University, Zhenjiang, Jiangsu 212013, PR China.  }
 \email{ksxu@ujs.edu.cn}

\affiliation{Centro Interdisciplinario de Neurociencia de Valpara\'iso, Universidad de Valpara\'iso, Valpara\'iso, 2360102, Chile
}%

\author{Jean Paul Maidana}
\author{Patricio Orio}
 \affiliation{Centro Interdisciplinario de Neurociencia de Valpara\'iso, Universidad de Valpara\'iso, Valpara\'iso, 2360102, Chile
}%

 \affiliation{Instituto de Neurociencia, Facultad de Ciencias, Universidad de Valpara\'iso, Valpara\'iso, 2360102, Chile..}
 \email{patricio.orio@uv.cl}


\date{Received: date / Accepted: date}


\begin{abstract}
Many experiments have evidenced that electrical and chemical synapses -- hybrid synapses -- coexist in most
organisms and brain structures. The role of electrical and chemical synapse connection in diversity of neural
activity generation has been investigated separately in networks of varying complexities. Nevertheless, 
theoretical understanding of hybrid synapses in diverse dynamical states of neural networks for self-organization and robustness still has not been fully studied.  Here, we present a model of neural network built with hybrid synapses to investigate the emergence of global and collective dynamics states. This neural networks consists of excitatory and inhibitory population interacting together. The excitatory population is connected by excitatory synapses in small
world topology and its adjacent neurons are also connected by gap junctions. The inhibitory population is only
connected by chemical inhibitory synapses with all-to-all interaction. Our numerical simulations  show that in the balanced 
networks with absence of electrical coupling, the synchrony states generated by this architecture are mainly
controlled by heterogeneity among neurons and the balance of its excitatory and inhibitory inputs. In balanced networks with strong electrical coupling, several dynamical states arise from different
combinations of excitatory and inhibitory weights. More importantly, we find that these states, such as synchronous firing, cluster synchrony,
and various ripples events, emerge by slight modification of chemical coupling
weights. For large enough electrical synapse coupling, the whole neural networks become synchronized.
Our results pave a  way in the study of the dynamical mechanisms and computational significance
of the contribution of mixed synapse in the neural functions. 
\end{abstract}

\maketitle

\section{Introduction}
\label{intro}
Over the past decades, spatiotemporal patterns  has been observed in experimental recordings of neural activity, such as synchronous states\cite{elson1998synchronous,jing2017network},  complex spatiotemporal patterns\cite{coombes2010large,muller2018cortical,ma2015wave,wang2007ordered,Xu_2013,xu2014simplified,de2011spatiotemporal} and chimera states\cite{tian2017diversity,huo2019chimera}, has long been of interest, but  their underlying mechanism of the generation of these patterns has not been fully understood yet.

It is generally accepted that there are  two main modalities of synaptic transmission -- chemical and electrical\cite{pereda2014electrical,alcami2019beyond}.  At chemical synapses, information is transferred through the release of a neurotransmitter from presynaptic neuron and detection of the neurotransmitter by an adjacent postsynaptic cell\cite{Sheng2012Synapse}, whereas in electrical synapses, the cytoplasm of adjacent cells is directly connected by clusters of intercellular channels called gap junctions\cite{bennett2004electrical}.  From this different characteristics and functionality, chemical synaptic communication is typically unidirectional and episodic.  By contrast,  electrical synaptic transmission through gap junctions underlies direct, continuous, reliable and rapid neuronal communication in the CNS\cite{bennett2004electrical}. In addition, electrical synapses are bidirectional and local connections between neurons which are spatially very close.   Theoretical work has shown that diversity of firing patterns and cortical functions is provided by balance between coactivated excitatory and inhibitory synaptic inputs\cite{anderson2000orientation,ATALLAH2009566,Wehr2003,Cindy2009Odor} (for a
review see ref.\cite{ISAACSON2011231}). For example, when externally driven, circuits of recurrently connected by excitatory and inhibitory neurons with strong synapses settle rapidly into a state in which population activity levels
ensure a balance of excitatory and inhibitory currents\cite{vanVreeswijk1724,Bittner2007Population,RubinE9366}. Over all,  excitatory neurons are thought to do only one thing: excite each other, whereas inhibitory neurons can generate nonlinear effects\cite{Buzsaki2006network}, such as modulating stimulus response gain, sharpening  tuning to stimuli, and pacing cortical oscillations\cite{ISAACSON2011231,Buzsaki2006network}.  For electrical coupling, it is responsible for a variety of network effects, particularly in networks that generate rhythmic activity, some of which are well established, such as regulation of
phase relationships, synchrony, and pattern formation\cite{beierlein2000network,bucher2005animal,Marder2007Understanding,NARGEOT2009975,Wenning2011}, and some are novel, such as a direct role in rhythmogenesis (also see review\cite{jing2017network}). 

The relationship between spatiotemporal patterns of neuronal activity and  the types of synaptic
transmission can be investigated in modeling studies\cite{bartos2007synaptic,roxin2005role,Ermentrout_1998,townsend2018detection}. 
Computational models have shown that the synchronous occurrence of action potentials in many neurons is mainly due to connections of chemical inhibitory synapses\cite{traub1999fast,buzsaki2012mechanisms,fisahn1998cholinergic,buzsaki1995temporal,benardo1997recruitment}  or electrical synapses between neurons\cite{elson1998synchronous,skinner1999bursting,sohl2005expression,tang2011synchronization,xu2018synchronization,jing2017network}.  
 The results obtained by B{\"o}rgers et al. \cite{borgers2003synchronization} also suggest that synchronous oscillations  often comes from interplay of chemical synapses connections between E-cells (excitatory neurons) and I-cells (inhibitory neurons) but under ideal conditions -- homogeneity in relevant network parameters and all-to-all-connections. Synchronous states have traditionally been thought to involve precise zero-lag synchrony\cite{gray1989oscillatory,uhlhaas2009neural}.
Instead of precise zero-lag synchrony, a range of flexible phase offsets -- phase differences between two (or more)
oscillations-- is possible. These flexible phase
relationships can, in their simplest form, be traveling waves of various shapes (including plane, radial and spiral
waves\cite{keane2015propagating}). The combination of multiple traveling waves can form complex spatiotemporal patterns \cite{muller2012propagating,muller2018cortical}. It has been shown, in spiking neural networks\cite{keane2015propagating,mehring2003activity,yger2011topologically,voges2012complex,senk2018conditions}that transmission delays\cite{tang2011synchronization}, the spatial reach of
connections\cite{Xu_2013} and the strength of inhibition of the excitatory-inhibitory networks influence the emergence of spatiotemporal patterns, such as asynchronous and irregular activity, or propagating waves. Most previous modeling studies have shown that the formation of spatiotemporal patterns  such as traveling waves in neural networks has mainly
been investigated by means of phenomenological neural-field models. These models uncover the existence and uniqueness of diverse patterns which are stationary or nonstationary in space and time, such as waves, wave fronts, bumps, pulses, and periodic patterns\cite{Ermentrout_1998,Coombes2005,Bressloff2014}. Recently, much attention has been paid to the chimera state\cite{martens2010bistable,omelchenko2011loss,tian2018chimera,martens2010solvable}.  In chimera state, populations of coupled oscillators may exhibit two coexisting subpopulations, one with synchronized oscillations and the other with unsynchronized oscillations, even though all of the oscillators are coupled to each other in an equivalent manner\cite{omel2008chimera,tinsley2012chimera}. This
symmetry-breaking behavior has been studied in a variety of complex dynamical systems and has attracted growing
attention in neural system studies from both theoretical and experimental perspectives\cite{hizanidis2016chimera,majhi2017chimera,calim2018chimera}.

The role of purely electrical synapse or chemical synapse connections in generation of the diversity of neural activity  has been investigated deeply in networks of varying complexities as discussed above. Many experiments have evidenced that diverse neural circuits use a combination of electrical and chemical synapses -- hybrid synapses -- coexisting in most organisms and brain structures\cite{rash2005ultrastructural,vivar2012mixed,hamzei2012mixed,kuo2016nonlinear},  to convey signals between neurons (reviewed in\cite{nagy2018electrical,pereda2014electrical} ). At first, the synchronous dynamical behavior of the neuronal systems with two types of synapses have attracted more and more attention. Kopell et al.\cite{kopell2004chemical} found that both electrical and chemical synapses play the different but complementary roles in the synchronization of neuronal networks. Baptista et al.\cite{baptista2010combined} further combined two types of synapses to study the process of synchronization and the information production. Liu et al.\cite{liu2013impact} studied the combined effects of the information transmission delay and the ratio of the electrical and chemical synapses on the synchronization transitions.  Secondly,  the concept in which chimera-like states emerge spontaneously with suitable tunning of electrical and chemical coupling strengths, surfaced in the pioneering theoretical  and  computational work of Hizanidis et al.\cite{hizanidis2016chimera}. In another work, Majhi et al.\cite{majhi2017chimera} analyzed chimera states in a two-layer neural network connected with hybrid synapses. The results obtained by them showed that the emergence of chimera states depends significantly on chemical synapses, not electrical ones. More recently,  Calim et al.\cite{calim2018chimera} highlighted the importance of the excitability level and nonlocal network features on spatiotemporal patterns in networks of neurons coupled  only via chemical synapses. Furthermore, they mapped the transitions between incoherent states, traveling waves, chimeras, coherent states, and global amplitude death in the parameter space. More importantly,  Pernelle et al.\cite{pernelle2018gap} showed that gap junction plasticity acts as a mechanism to regulate oscillations in  the  networks connected by hybrid synapses.

All these approaches mentioned above only  identify the essential mechanisms for generating various complex spatiotemporal patterns (such as synchrony state, chimeras state, and etc) in neural networks coupled via chemical synapse or hybrid synapses. Nevertheless, they give little insight on how the electrical and chemical synapses are related in the emergence of a diversity of dynamical states in neural networks for self-organization and robustness. 
In other words, the relationship between spatiotemporal patterns of neuronal activity and the types of synaptic transmission together with corresponding synaptic weights has not been systematically investigated yet. For example, whether the excitatory-inhibitory  balance or gap junction make significant different contributions to generate firing patterns , and whether these different states can coexist in the same  parameter space or the same time bins of interest still have not been fully studied yet. 

Our first  finding is that, in the E/I balanced networks of excitatory population in absence of gap junction, the excitatory population has  a tendency to cluster synchronization as the weight of inhibitory synaptic currents into the excitatory part is increase. In other words,  the larger the chemical inhibitory synapse strength used for excitatory part the higher degree of synchronization to achieve. However, introducing the gap junction to the excitatory population, it usually can go through several phases of network state, that are asynchronous activity, synchronous firing,  various ripples events  (such as traveling wave) and lastly  subthreshold oscillations without spikes. For example, we find that  for  stronger inhibitory synaptic strengths larger than a certain amount  leads to suppression of the spikes oscillation, and for moderate  inhibitory synaptic strengths one should expect to find chimera-likeness behaviors, which is also metastable state in some time windows.  More importantly, it first time to show the coexistence of these dynamics states, such as coexistence of coherent state, traveling wave and oscillation without spikes,  depending on the inhibition level . Finally, our results suggest  that if  excitatory populations are connected with hybrid synapses,  the  excitation and inhibition balance can  make a great contribution to generate various spatiotemporal patterns.

\section{Models and methods}

\subsection{Neuronal dynamics}

\textbf{Wang-Buzsáki model.} The Wang-Buzsáki model we used has been presented detailly in Ref\cite{wang1996gamma,palmigiano2017flexible}, here we give a brief overview. This conductance-based model has three states variables for each neurons: membrane potential of
each cells $V$, the Sodium inactivation variable $h$ and Potassium activation variable $n$ corresponding to the spikes-generating $Na^+ $ and $K^+$ voltage-dependent ion currents ($I_{Na}$ and $I_K$), respectively. Sodium activation variable $m$ is considered to be instantaneous. The neuronal dynamics can be described as :
\begin{align}
	C_m\frac{dV}{dt} &=-I_{Na}-I_{K}-I_{L}-I_{Syn}+I_{app} \nonumber \\
	\frac{dh}{dt} &= \phi(\alpha_h(1-h)-\beta_hh) \nonumber \\
	\frac{dn}{dt} &= \phi(\alpha_n(1-n)-\beta_nn) 
	\end{align}
	
 Where $I_L  = g_L(V-E_L)$, $I_{Na} = g_{Na}m^{3}_{\infty}h(V-E_{Na})$  and  $I_K = g_K n^{4}(V-E_K)$ represent the leak currents, transient sodium currents and the delayed rectifier currents, respectively. $I_{syn}$ stands for the synaptic currents and $I_{app}$ is the injected currents (in $\mu A/cm^2$). The parameters $g_L$, $g_{Na}$, $g_K$ are the maximal conductance density, $E_L$, $E_{Na}$, $E_K$ are the reversal potential and function $m_{\infty}$ is the steady-state activation variables $m$ of the Hodgkin-Huxley type\cite{hodgkin1952quantitative}.

\begin{figure}[H] 	
	\includegraphics[width=\linewidth]{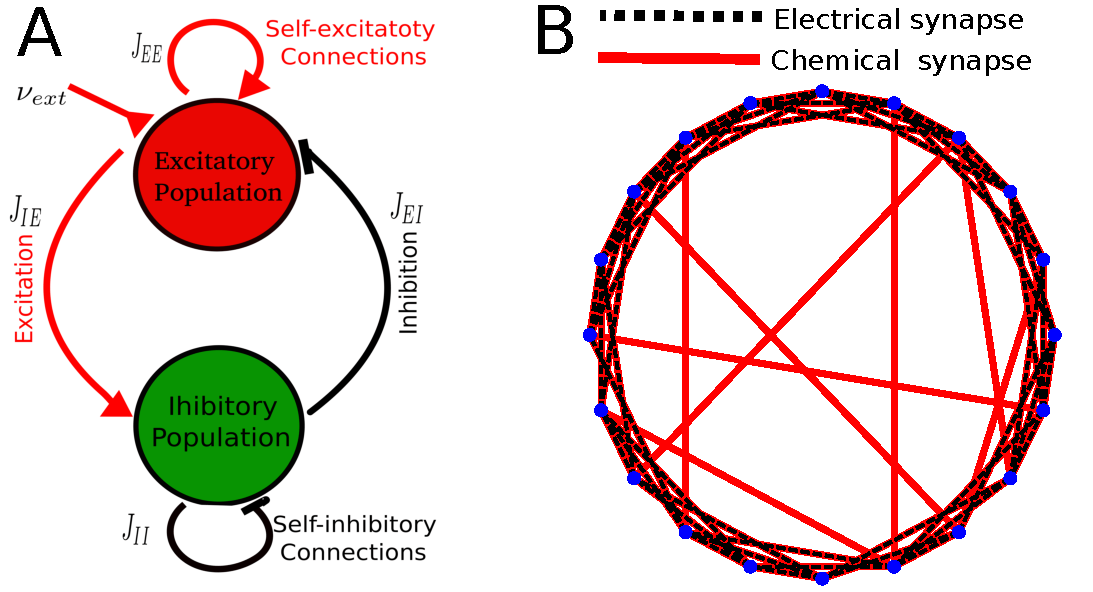}	
	\caption{ Network architecture and parameters. \textbf{A}. The balanced neural networks containing excitatory ({\it red}) and inhibitory populations ({\it green}). The synaptic input (excitatory synaptic input or inhibitory synaptic input) between the populations are generated by all-to-all connections. Their coupling strength between the populations is denoted with constants $J_{IE}$ and $J_{EI}$. Self-inhibitory connections (of the weights $J_{II}$) is all to all connections.  Excitatory (inhibitory) connections are represented in red ({\it black}). $\nu_{ext}$ is the external drive to each neuron in the excitatory population by external excitatory inputs. \textbf{B}. Diagram of excitatory populations connected by electrical (linear diffusive coupling) and chemical (nonlinear coupling) synapses. The excitatory population is connected by excitatory synaptic  ({\it red solid lines}) strength $J_{EE}$ in a small world topology and adjacent neurons are also
connected by electrical synaptic strength $J_{gap}$ ({\it black dashed lines} ). Without loss of generality, we fix $J_{II} = 0.04$, $J_{IE} = 0.01$. }
	
	\label{fig:Schematic_figure}
\end{figure}

   In the simulation, unless stated otherwise, the parameters and functions used are given in Table \ref{S1_Table}.

\begin{widetext}
\begin{table}[H] 
\caption{ Parameters and functions of Wang-Buzsáki model(1996)\cite{wang1996gamma}}
	{\renewcommand{\arraystretch}{1.4}%
	\begin{tabular}{ p{5cm} p{5cm} p{5cm}  }
	    \hline \\ 
	    $g_{L}=0.1 $ & $g_{Na}=35$ &$g_{K}=9$ ($mS/cm^2$)\\
		$E_{L}=-65$   & $E_{Na}=55$   &$V_{K}=-90$ (mV)\\
		$C_m=1$ ($\mu F)$ &  $\phi = 5$   &  $I_{app}  = 0$ ($\mu A/cm^2$)\\
	\end{tabular}
	
	\begin{tabular}{p{9.5cm} p{5.5cm}}
		$\alpha_{m}(V)=-0.1(V+35)/[exp(-0.1(V+28))+1]$ & $\beta_{m}(V)=4exp[-(V+60)/18]$ \\
		$\alpha_{h}(V)=0.07exp[-(V+58)/20]$ & $\beta_{h}(V)=1/exp[-0.1(V+28)]+1$ \\	
		$\alpha_n(V) = -0.01(V+34)/[exp(-0.1(V+34))-1]$ & $\beta_n(V) = 0.125exp[-(V+44)/80]$ \\
		$ m_{\infty}=\alpha_{m}/(\alpha_{m}+\beta_{m})$ &  $ $  \\
		 &  \\
		\hline
	\end{tabular}}

\label{S1_Table}
\end{table} 
\end{widetext}

\textbf{Synaptic dynamics.} By applying the Wang-Buzsáki model to the nodes of the network that incorporates both types of synapses, neurons in the networks are connected by electrical (linear diffusive coupling) and chemical (nonlinear coupling) synapses. The total synaptic input currents into neuron  $i$ within excitatory population, is given by:

\begin{widetext}
\begin{align}\label{eq:excSynaCurrents}
	I_{SynE,i} &= J_{gap}\sum_{j=1,j\ne i}^{N_E}C_{gap}^{ij}(V_{i}-V_{j}) + J_{EE}\sum_{j=1,j\ne i}^{N_{E}}L_{EE}^{ij}g_{syn,i}(V_{i}-E_{synE}) \nonumber \\
& +  J_{EI}\sum_{j=1,j\ne i}^{N_{I}}L_{EI}^{ij}g_{syn,i}(V_{i}-E_{synI})+ \sum_{i=1}^{N_{E}}g_{syn,i}(V_{i}-E_{synE})
\end{align}
\end{widetext}

The four terms on right-hand side of  \hyperref[eq:excSynaCurrents]{Eq.(\ref{eq:excSynaCurrents})}  corresponds 
to electrical synaptic inputs, self-excitatory synaptic inputs, inhibitory synaptic inputs (from  inhibitory population) and external excitatory synaptic input to each neuron in the excitatory population, respectively. The connection matrices $C_{gap}^{ij},L_{EE}^{ij},L_{EI}^{ij}$  are described in the network connectivity section.  Lastly   purely chemical synaptic input currents  for inhibitory population is as follows:

\begin{widetext}
\begin{align}\label{eq:inhiSyna}
	I_{SynI,i} =  J_{II}\sum_{j=1,j\ne i}^{N_{I}}L_{II}^{ij}g_{syn,i}(V_{i}-E_{synI}) + J_{IE}\sum_{j=1,j\ne i}^{N_{E}}L_{IE}^{ij}g_{syn,i}(V_{i}-E_{synE})
\end{align}
\end{widetext}

The two terms on right-hand side of \hyperref[eq:inhiSyna]{Eq.(\ref{eq:inhiSyna})} represent self-inhibitory synaptic input and excitatory synaptic inputs (from  excitatory population).

 We use a second-order differential equation to model chemical synaptic conductances $g_{syn}$\cite{koch1998methods,sterratt2011principles}, provided by:
 
\begin{widetext}
\begin{align}
\frac{d^2g_{syn}}{dt^2}={\bar{g}}_{s(E,I,\nu)}f\delta\left(t_{0}+t_{d}-t\right)-\left(\frac{1}{\tau_1}+\frac{1}{\tau_2}\right)\frac{dg_{syn}}{dt} -\frac{g_{syn}}{\tau_1\tau_2} 
\end{align}
\end{widetext}

The  delta function $\delta\left(t_{0}+t_{d}-t\right)$ represents the spike signal from presynatpic cell. The time $t_{0}$ is the time of the presynaptic spike and $t_{d}$ is synaptic delays from the presynatpic to postsynaptic cell.  The conductance peaks occurs at time $t_{peak} = t_{0}+t_{d}+\frac{\tau_{1}\tau_{2}}{\tau_{1}-\tau_{2}}ln(\tau_{1}/\tau_{2})$. The normalization factor $f$, ensures that the amplitude equals ${\bar{g}}_{s(E,I,\nu)}$, follows:
 
\begin{widetext}
\begin{align}
f=\dfrac{1}{e^{-\left(t_{peak}-t_0-t_d\right)/\tau_{1}}-e^{-\left(t_{peak}-t_0-t_d\right)/\tau_{2}}}
\end{align}
\end{widetext}

In simulations and for practical reasons, second order differential equations are solved as follows:
 
\begin{widetext}
\begin{align}
\frac{dy}{dt}&={\bar{g}}_{s(E,I,\nu)}f\delta\left(t_{0}+t_{d}-t\right)-\left(\frac{1}{\tau_1}+\frac{1}{\tau_2}\right)y -\frac{g_{syn}}{\tau_1\tau_2} \nonumber \\
\frac{dg_{syn}}{dt}&=y  
\end{align}
\end{widetext}

In simulations, both peak synaptic conductances ${\bar{g}}_{s(E,I,\nu)}$ and synaptic delays $t_{d}$ are Gaussian distributed random variables with prescribed means $g_{s(E,I)}$ and $\overline{D}$, and standard deviation $\sigma_{g_{s(E,I)}}$ and $\sigma_{\overline{D}}$. The background drive $v_{ext}$ to each neuron in the excitatory population was provided by external excitatory inputs, modeled as independent and identically distributed Poisson processes with rate $\nu$ for different neurons. Peak conductances of the external drive were also heterogeneous and assumed values sampled from a truncated Gaussian distribution with mean $g_{s\nu}$ and standard deviation  $\sigma_{g_{s\nu}}$.  These rules are the same as the previous study\cite{palmigiano2017flexible}. Parameters of synaptic model are summarized in  Table  \ref{S2_Table}.

\begin{widetext}

\begin{table}[H] 
	{\renewcommand{\arraystretch}{1.4}%
	\begin{tabular}{ |p{10cm}| p{2.5cm}| p{2.5cm}|  }
	    \hline
	     \rowcolor{black!80} \multicolumn{3}{l}{\textcolor{white}{A: Global simulation paramters}}  \\ \hline
	    
	    \textbf{Description} & \textbf{Symbol} & \textbf{Value}\\ \hline
	    Simulation duration &  $T_{sim}$ & 2000 ms \\ \hline
	    Start-up transient & $T_{trans}$  & 400 ms \\ \hline
	    Time step  &  dt &    0.02 ms\\ \hline
	    
	    \rowcolor{black!80} \multicolumn{3}{l}{\textcolor{white}{B: Populations and external input}}  \\ \hline
	    \textbf{Description} & \textbf{Symbol} & \textbf{Value}\\ \hline
	    Population size of excitatory neurons& $N_{E}$   & 1000 \\ \hline
	    Population size of inhibitory neurons  & $N_{I}$   & 250 \\ \hline
	    Possion input rate (external excitatory inputs)  &  $\nu$ & 6000 Hz \\ \hline
	    
	    \rowcolor{black!80} \multicolumn{3}{l}{\textcolor{white}{C: Connection parameters}}  \\ \hline
	    \textbf{Description} & \textbf{Symbol} & \textbf{Value}\\ \hline
		Weight of gap junction   & $J_{gap}$   & 0.1 \\ \hline
		Weight of self-excitatory connection   & $J_{EE}$   & 0.5 \\ \hline
		Weight of self-inhibitory connection   & $J_{II}$   & 0.04 \\ \hline
		Weight of inhibitory connections for excitatory populations  & $J_{EI}$   & 0.03 \\ \hline
        Weight of excitatory connections for inhibitory populations  & $J_{IE}$   & 0.01 \\ \hline
        Probability of local inhibitory connections &  $P_{I}$ & 0.2  \\ \hline
        
		\rowcolor{gray!60} \multicolumn{3}{c}{Synaptic dynamics (Difference of Two Exponentials)}  \\ \hline
		Reversal potential for excitatory synapses   & $E_{synE}$   & 0 mV \\ \hline
		Reversal potential for inhibitory synapses   & $E_{synI}$   & -80 mV \\ \hline
		Excitatory synaptic decay time   & $\tau_{1E}$   & 3 ms \\ \hline
		Ihibitory synaptic decay time   & $\tau_{1I}$   & 4 ms \\ \hline
		Excitatory synaptic time & $\tau_{2E}$&  1 ms \\ \hline
		Inhibitory synaptic time & $\tau_{2I}$&  1 ms \\ \hline
		Mean synaptic delay &  $\overline{D}$ &  1.5 ms \\ \hline
		Mean synaptic excitatory conductance &  $g_{sE}$ &  5 nS \\ \hline
		Mean synaptic  inhibitory conductance & $g_{sI}$ & 200 nS \\ \hline
		Mean synaptic  input conductance & $g_{s\nu}$ & 3 nS \\ \hline
		Standard deviation delay & $\sigma_{\overline{D}}$& 0.1 ms \\ \hline
		Standard synaptic excitatory conductance & $\sigma_{g_{sE}}$& 1 nS\\ \hline
        Standard inhibitory conductance & $\sigma_{g_{sI}}$& 10 nS \\ \hline
        Standard input conductance & $\sigma_{g_{s\nu}}$& 1 nS \\ \hline
        		
	    \rowcolor{gray!60} \multicolumn{3}{c}{Small-world networks (SW)}  \\ \hline
	    Probability of replacing new links   &  $P_{sw}$ & 0.01  \\ \hline
		Degree of nearest neighbours connections    & K & 10 \\ 
		\hline
	\end{tabular}}
	\caption{ List of simulation and default network parameters in the E/I balanced networks.  }
\label{S2_Table}
\end{table}
\end{widetext}

\subsection{Network connectivity}
\label{Net_matrix}
To build the connectivity matrix, we consider a balanced neural network, consisting of $N_E$ excitatory  and  $N_I$ inhibitory neurons shown in \hyperref[fig:Schematic_figure]{Fig.\ref{fig:Schematic_figure}}.  We will use the subscript $E$  to denote the excitatory populations and $I$  for the inhibitory ones.  The connections between the $i$th neurons of the $kth$ population and the $jth$ neurons of the $lth$ population, denoted by $L_{kl}^{ij}$. Here $k,l = E, I$. The neurons in excitatory population are connected by chemical synapses in a small world topology and adjacent neurons are  connected by gap junctions. The connection matrices of chemical and electrical synapses in the excitatory populations are defined as $L_{EE}^{ij}$ and $C_{gap}^{ij}$, while the connection matrices of chemical synapses inside the inhibitory one are given by $L_{II}^{ij}$. The weights of self-excitatory connections or self-inhibitory ones is symbolized by $J_{EE}$ or $J_{II}$. The coupling strengths of adjacent electrical connections are denoted with $J_{gap}$.  The link matrices between populations with all-to-all connections are $L_{IE}^{ij}$ and $L_{EI}^{ij}$. Their coupling strength between the populations is denoted with constants $J_{IE}$ and $J_{EI}$. The connectivity  matrix $L_{kl}^{ij} = 1$ ($L_{kl}^{ij} = 0$) corresponds to connections between neuron $i$ and $j$ (or not).  The  default network parameters used  are shown in Table \ref{S2_Table}.

 The small-world topology mentioned above  is implemented as two basic steps of the standard algorithm\cite{Watts1998}: 1. Construct a regular ring lattice with $N$ nodes of mean degree $2K$. Each node is connected to its $K$ nearest neighbors on either side. 2. For every node $i=0,..., N-1$ take every edge connecting $i$ to its $K$ rightmost neighbors -- that is every edge ($i,j$ mod $N $) with $i<j \leq i+K$ , and rewire it with probability $P_{sw}$. Rewiring is done by replacing ($i,j$ mod $N $) with ($i,k$) where $k$ is chosen uniformly at random from all possible nodes . Self-inhibitory connections is all-to-all connections. The rewiring edge cannot be a duplicate or self-loop.
\subsection{Network Activity Characterization}
  In this paper we first use the \textbf{synchronization index $\chi(N)$}, previously introduced by Ref \cite{golomb1993dynamics,golomb1994clustering} ,  to account for the synchronization level of the neural activity of the considered networks, where:

\begin{align}\label{eq:definition_index}
\chi^2=\dfrac{N\sigma_{V(t)}^2}{\sum_{i=1}^N\sigma_{V_{i}(t)}^2}
\end{align}
Here $V(t) = \frac{1}{N}\sum_{i=1}^NV_{i}(t) $ is populations average of the membrane potential $V_{i}(t)$. The variables $\sigma_{V(t)}$ and $\sigma_{V_{i}(t)}$ denote the standard deviation of $V(t)$ over time, or respectively, of the membrane potential traces $V_{i}(t)$ of the each isolated neuron $i$.

To further investigate the global dynamical behaviour of the neural networks, we then introduce the \textbf{order parameter} and  its the variance in time, \textbf{metastability}. The order parameter $R$, describes the global level of phase synchrony in a system of $N_E$ oscillators\cite{kuramoto2003chemical,bertolotti2017synchronization}, given by: 

\begin{equation} \label{eq:definition_R}
R= \left\langle \left| \frac{1}{N_E} \sum_{k=1}^{N_E} e^{2\pi i[(t-t_k^n)/(t_k^{n+1}-t_k^n)]}  \right| \right\rangle
\end{equation}

Where  the vertical bars ($=\phi_c(t) $) denotes the modulus of the complex
number and the angle brackets is the temporal average value of that quantity, while the exponent, $2\pi[(t-t_k^n)/(t_k^{n+1}-t_k^n)]$, is the phase  we assign to each neurons in  the excitatory  populations. The closer to 0 (1) $R$ becomes, the more
asynchronous (synchronous) the dynamics is. $N_E$ is the total number of excitatory populations, $t_k^n$ is the time of the $n$th spikes of neurons $k$ and $t_k^{n+1}$ is the time of the following $(n+1)$th spikes.

In order to quantify the metastability and chimera-likeness of the observed dynamics, the global metastability $Met$\cite{shanahan2010metastable} , is the variance in time of  order parameter $R$, by:

\begin{equation}\label{eq:metastability}
Met= \frac{1}{\Delta t } \sum \limits_{t\leq\Delta t}\left(\phi_c(t)-R\right)^2
\end{equation}

where $\Delta t$ in \hyperref[eq:metastability]{Eq.(\ref{eq:metastability})} is the time windows to quantify  the metastability of the excitatory populations. Metastability is $0$ if the system is either completely synchronized or completely desynchronized -- a high value is present only when periods of coherence alternate with periods of incoherence.

\section{Results}
\subsection*{\textbf{Influence of the electrical coupling in the generation of several firing patterns }}

In what follows, our main goal is to investigate how the electrical and chemical coupling  can affect the  firing patterns diversity within the excitatory population of the E/I balanced networks. To do this, we simulated  balanced networks  containing  excitatory and inhibitory populations.  Neurons in the excitatory population are connected by both electrical and chemical synapses, while  neurons in inhibitory one are only coupled by chemical synapses (See \ref{Net_matrix} section  for detail).  In this work, we consider the network composed of Wang-Buzsáki neurons. To measure the level of synchronization,  we use the synchronization index $\chi(N)$ (\hyperref[eq:definition_index]{Eq.(\ref{eq:definition_index})}) and Kuramoto order parameter $R$ (\hyperref[eq:definition_R]{Eq.(\ref{eq:definition_R})}).  We also introduce the metastability $Met$  \hyperref[eq:metastability]{Eq.(\ref{eq:metastability})} to quantify the metastability and chimera-likeness of the observed dynamics within excitatory populations.

\begin{figure}[H] 	
	\includegraphics[width=\linewidth]{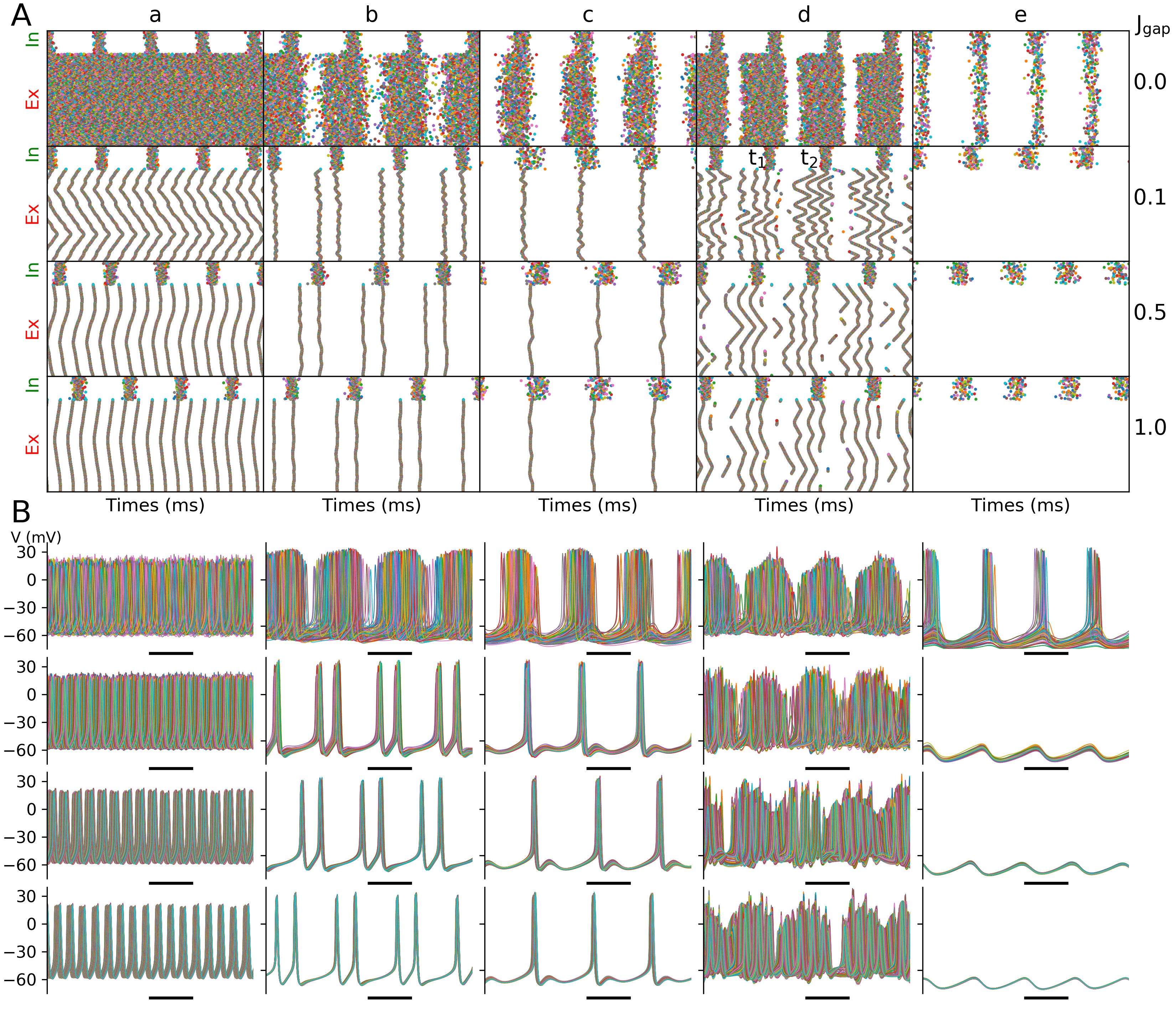}	
	\caption{ Examples of spatiotemporal firing patterns (\textbf{A}) and  their corresponding action potentials (\textbf{B})  in each dynamical regimes  against several  level of  electrical strengths $J_{gap}$ in ($J_{EE},J_{EI}$) parameter space.  The 1$th$ row and 2-4$th$ rows in (\textbf{A}) correspond the excitatory population built with or without gap junction, respectively.  Raster plots of spatiotemporal firing patterns within 5 different dynamical regimes (a-e) that are shown in  \hyperref[fig:ParaSwape_JeeVsJei]{Fig.\ref{fig:ParaSwape_JeeVsJei}}. Example of chimera-likeness  and incoherent state  respectively at time $t_1$ and $t_2$ in point $d$ with $J_{gap}=0.1$. Metastable state is that synchronous part of
chimera-likeness state alternates between synchronous and incoherent behaviors with different times, eg. $t_1$ and $t_2$. Weights of gap junction $J_{gap}$ from top to bottom panels are $0.0$, $0.1$, $0.5$, $1.0$. The red  Ex (green In)  of  $y$ labels in  (\textbf{A}) indicates the neurons index of excitatory population (inhibitory population).   In all traces the scale bar in (\textbf{B}) is 10 ms.  }
	
	\label{fig:Figure3_rasterplot}
\end{figure}

\begin{figure}[H] 	
	\includegraphics[width=\linewidth]{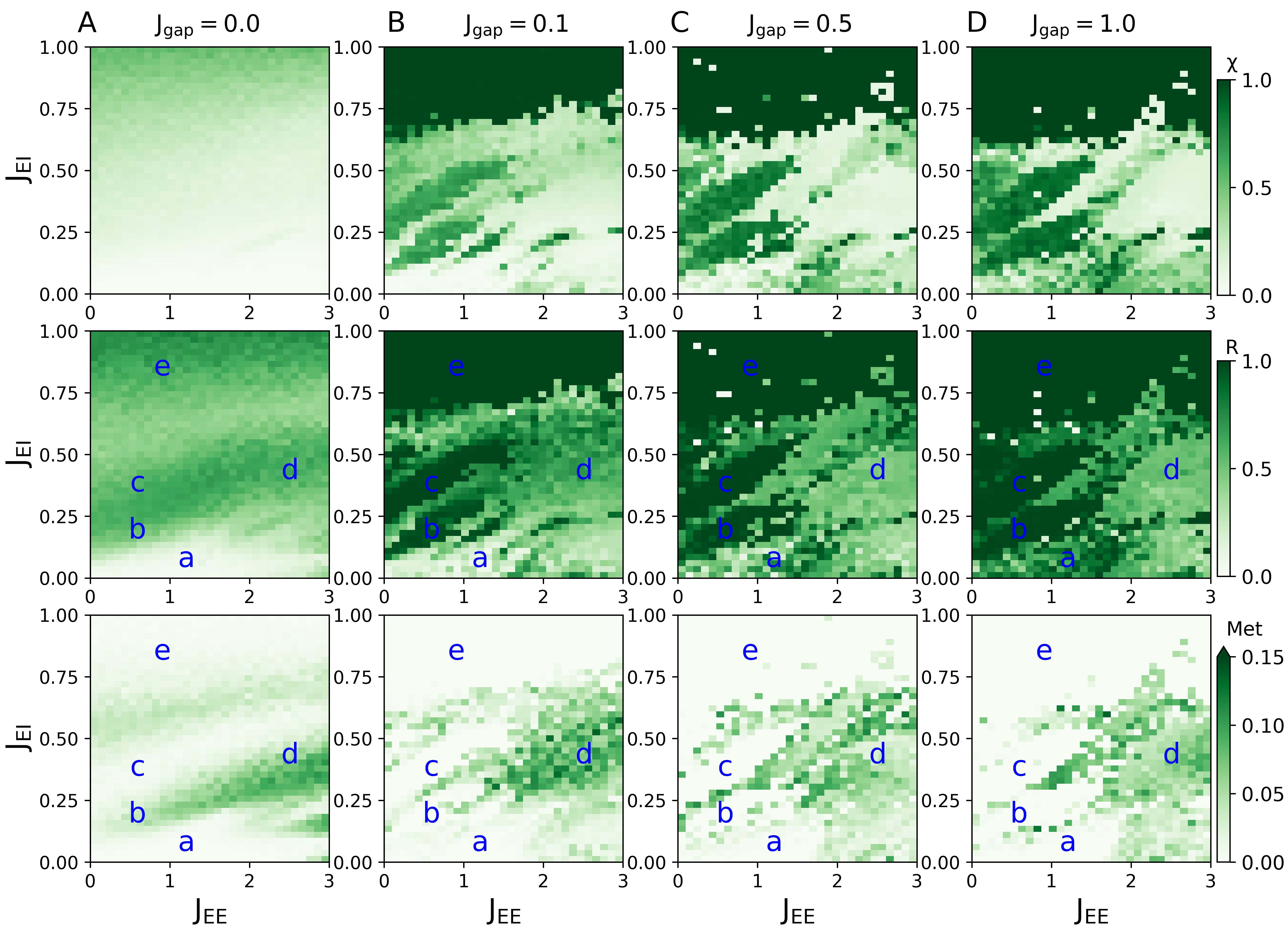}	
	\caption{ Dynamical regimes of spatiotemporal firing patterns on ($J_{EE},J_{EI}$) parameter space with different weights $J_{gap}$ of electrical synaptic connections. Density plots show synchronization index $\chi(N)$ (\textbf{Upper panels}) as well as order parameter $R$ (\textbf{Middle panels}) and metastability $Met$ (\textbf{Bottom panels}) of excitatory populations in the same plane. The marked points $a,b,c,d,e$  denote  different dynamical regimes.
	\textbf{A}.Collective dynamics behavior of excitatory populations with purely chemical synaptic connections ($J_{gap}=0$).  (a-c) Incoherent state corresponds to low values of both, $\chi$ and $Met$; (d)Metastable state  where $\chi$ is low and $Met$ is high; (e) Generalized synchronous state (not fully synchronous state) where $\chi$ is  much higher but less than 1 and $Met$ is low.  Subplots \textbf{B}-\textbf{D}  illustrate different dynamics regimes in network of excitatory population built with hybrid
synapses: (a) Traveling wave (wave propagation) where both $\chi$ and $Met$ are low values; (b) and (c) Synchronous states;  (d) Metastable state or chimera-likeness where $\chi$ is low and $Met$ is high; This indicates a highly metastable behavior and desynchronous dynamics; (e) Subthreshold synchronous states (but neurons within excitatory populations oscillate without spikes).   The electrical weights  for each subplot  are  $J_{gap}=0.1$, $J_{gap}=0.5$, $J_{gap}=1$, respectively.}
	
	\label{fig:ParaSwape_JeeVsJei}
\end{figure}

\hyperref[fig:Figure3_rasterplot]{Fig.\ref{fig:Figure3_rasterplot}} shows the typical examples of spatiotemporal firing patterns (\textbf{A}) and corresponding action potentials (\textbf{B}) from  each dynamical regime (corresponding marked by letters $a,b,c,d,e$ in \hyperref[fig:ParaSwape_JeeVsJei]{Fig.\ref{fig:ParaSwape_JeeVsJei}}), and their changes with different degree of electrical synapse connections. If excitatory  population of  networks  are only  built  with  the chemical synaptic  connections,  raster plots (\textbf{Row} 1  in \hyperref[fig:Figure3_rasterplot]{Fig.\ref{fig:Figure3_rasterplot}A}) show that  the larger the inhibitory  strengths used  for excitatory population the  higher level of synchronization to achieve,  and denote three different dynamical regimes -- incoherent state, metastable state and generalized synchronous state (or cluster synchronous states), respectively. Incoherent state (\textbf{Columns} $a$ to $c$ in \textbf{Row} 1 of \hyperref[fig:Figure3_rasterplot]{Fig.\ref{fig:Figure3_rasterplot}})  is clear  in the network of excitatory population and  note that, firing patterns of panels $b,c$  easily tend to clustering. An interesting behavior is metastable state (\textbf{Column} $d$ of \hyperref[fig:Figure3_rasterplot]{Fig.\ref{fig:Figure3_rasterplot}}), but it is difficult to show in this case. The panel $e$ displays generalized synchronous state, this is because of stronger inhibition. The action potentials corresponding to these raster plots in excitatory population are plotted in the first panels of \hyperref[fig:Figure3_rasterplot]{Fig.\ref{fig:Figure3_rasterplot}B}. 

In the case of networks built with hybrid synapses, it is clear, from the raster plot of column $a$  with $J_{gap} =0.1$ (\textbf{Row} 2  of \hyperref[fig:Figure3_rasterplot]{Fig.\ref{fig:Figure3_rasterplot}A}), to find that there is an emergence of traveling wave (wave propagation), which can not be observed in the same parameter space without gap junction (see \textbf{Row} 1 of \textbf{Column} $a$).  On the other sides,  traveling waves observed  in excitatory populaton has a tendency to  synchrony with increasing the weights of gap junction (see \textbf{Rows} 2,3,4 of  \textbf{Column} $a$ with $J_{gap} = 0.1,0.5,1$,respectively). Columns $b,c,e$  (illustrated in \textbf{Row} 2  of \hyperref[fig:Figure3_rasterplot]{Fig.\ref{fig:Figure3_rasterplot}A} with $J_{gap} = 0.1$)  correspond to dynamical regimes with almost completely synchrony state, and the corresponding time series exhibits spiking behavior shown in \hyperref[fig:Figure3_rasterplot]{Fig.\ref{fig:Figure3_rasterplot}B}.
Note that, we have shown that raster plots and action potentials provide  totally discrimination in the observed synchrony regimes. The type of oscillatory synchronization  occurs in columns $b,c$ (\textbf{Row} 2) with suitable inhibition corresponding to the  oscillations with period 2 and 1, respectively, while column $e$ is related to  subthreshold oscillatory synchronization without spikes under  stronger inhibition. The result in  column $e$  shows that the cooperation of  electrical synapse connections and stronger inhibition leads to suppression of the spikes oscillation, namely subthreshold synchronized oscillations.    More importantly, increasing the value of electrical synapse connections weights causes no obvious change in the modes of synchronized oscillation (see  \textbf{Columns}  $b,c, e $ with $J_{gap} = 0.5,1$ in \hyperref[fig:Figure3_rasterplot]{Fig.\ref{fig:Figure3_rasterplot}A}).  An interesting highly metastable behavior is found in the  regime with higher inhibitory level, where $d$ is located in   \hyperref[fig:ParaSwape_JeeVsJei]{Fig.\ref{fig:ParaSwape_JeeVsJei}}. Raster plot of this point (in \textbf{Row} 2 of \textbf{Column} d  with $J_{gap} = 0.1$) displays chimera-likeness state, which coexists two subpopulations of synchronous states and incoherent state with various ripples events (such as traveling waves) in the same certain time, e.g. at time $t_1$. More importantly, chimera-likeness state sometimes  is also a metastable state in time, meaning that they can stay in the vicinity of one stable state for a certain time interval  and then, spontaneously move towards another-- such  as dynamics switching between time $t_1$ and $t_{2}$ shown in 
\hyperref[fig:Figure3_rasterplot]{Fig.\ref{fig:Figure3_rasterplot}A} (see also \hyperref[fig:Figure5_FixedJe25]{Fig.\ref{fig:Figure5_FixedJe25}}).  The firing patterns located in column $d$ have a slightly tendencty to synchrony as electrical connections are increase.  From the corresponding actions potentials  shown in  \hyperref[fig:Figure3_rasterplot]{Fig.\ref{fig:Figure3_rasterplot}B}, it is evident that the system is in the bursting regime. These results imply that the level and balance of  excitation and inhibition, when introducing the gap junction to the excitatory population, leads to various firing patterns, such as wave propagation, synchronized oscillations,  chimera-likeness and metastable state.

The dynamics of spatiotemporal firing patterns, in the excitatory populations with 4 different level of electrical coupling weights, are characterized in \hyperref[fig:ParaSwape_JeeVsJei]{Fig.\ref{fig:ParaSwape_JeeVsJei}} of the ($J_{EE}, J_{EI}$) parameter spaces.  These dynamical regimes are quantified via synchronization index as well as order parameter  and metastability.
In order to highlight some characteristic patterns that emerge in our system, we select 5 dynamical regimes of interest, marked by letters a, b, c, d and e in parameter spaces. Excitatory population network  built with hybrid synapses, are chosen so that the following cases are covered (see \hyperref[fig:ParaSwape_JeeVsJei]{Fig.\ref{fig:ParaSwape_JeeVsJei}}B-D). a: both  $\chi$ (or $R$  )  and $Met$ are low-valued; b,c and e:  $\chi$ is high-value but $Met$  is low-value; d:  $\chi$ is middle-value and $Met$  is high-value. To  further understand the role of electrical coupling on emergence of the diversity firing patterns,  the corresponding 5 points also  have been selected in the same parameter space of  excitatory population but without gap junction shown in \hyperref[fig:ParaSwape_JeeVsJei]{Fig.\ref{fig:ParaSwape_JeeVsJei}A} .

 We first investigate collective dynamical behavior of excitatory population with purely chemical synaptic connections ($J_{gap}=0$) in \hyperref[fig:ParaSwape_JeeVsJei]{Fig.\ref{fig:ParaSwape_JeeVsJei}A}. The results obtained in this figure display a region $e$ of lower level synchronization for high inhibitory chemical coupling with the rest of dynamical regimes being incoherent.  This is because the level of recurrent inhibition controls the overall level of synchrony of the excitatory population,  which is the first known reason to induce networks synchronization. However,  when networks of excitatory population built with hybrid synapses, with weaker electrical coupling shown in \hyperref[fig:ParaSwape_JeeVsJei]{Fig.\ref{fig:ParaSwape_JeeVsJei}B},  it exhibits  higher level synchronization ($\chi$ or $R$  is close to 1) than those of networks without gap junction in the same region $e$  for stronger inhibition. In addition, there are other two emergence regions ($b,c$) of high-level synchronzaiton where their synchronization index (or order parameter) are close to 1  for weaker inhibition. These two regions  can not be observed clearly in networks without electrical synapse connections (see corresponding regions $b,c$ in \hyperref[fig:ParaSwape_JeeVsJei]{Fig.\ref{fig:ParaSwape_JeeVsJei}A}). Apart from that, the contribution of electrical coupling on synchronization is also reflected in the same parameter space,  but with slight stronger weights of gap junction when $J_{gap} = 0.5, 1$ (Fig.\ref{fig:ParaSwape_JeeVsJei}C-D).   Note that,  increasing electrical synapse weights of excitatory population in network can enlarge regimes $b,c,d,e$ and shrink regime $a$,  comparing \hyperref[fig:ParaSwape_JeeVsJei]{Fig.\ref{fig:ParaSwape_JeeVsJei}B},C and D. This is the second mainly mechanism to increase the level of synchronization in a way that enhancing the electrical synapse  weights.  Besides,  it is also easy to see that the boundaries between each regime are obvious.

 Another usefully measurement related to network activity is the metastability, which measures the chimera-likeness/metastable state in the excitatory population along time. More surperisely, we found that introducing electrical connections, even with weaker coupling weights,  to excitatory population leads to  chimera-likeness/metastable in  region $d$ of the observed dynamics, as captured by metastability (bottom row of \hyperref[fig:ParaSwape_JeeVsJei]{Fig.\ref{fig:ParaSwape_JeeVsJei}B-D}). More importantly, enhancing weights of the electrical connections can  expand the metastability region $d$. It is clear that the region in which $Met$ attains values close to 0 coincides with all the neurons where excitatory population are in a synchronized state (region $b,c,e$) or incoherent state (region $a$). In other parts of the parameter space (region d) where $Met$ achieves higher values, indicating that the system is in  chimera-likeness state, coexisting two subpopulations of synchronous states and incoherent state,   or metastable state, often switching between synchronous and asynchronous states. The results shown above imply that in excitatory networks build with hybrid synapses, diverse dynamical regimes arises from different combinations of excitatory and inhibitory weights, and their boundary between each dynamical regime are clear, which can not display in the network  coupled by purely chemical synapses (bottom panel of \hyperref[fig:ParaSwape_JeeVsJei]{Fig.\ref{fig:ParaSwape_JeeVsJei}A}). 
 
Next we focus on the  influence of excitation and inhibition together with their balance to network firing patterns, in the networks of excitatory population built with hybrid synapses.

\subsection*{\textbf{Influence of  excitation and inhibition on excitatory population behavior}}
 We investigate now the emergence of various firing patterns with variation of inhibitory level accessible by the weights parameter $J_{EI}$. To do this,  we separately choose excitatory level $J_{EE} = 0.5$ and $2.5$,  in the networks of excitatory population built with ($J_{gap} = 0.1$) or without ($J_{gap} = 0$) electrical synapse connections.

\begin{figure}[H] 	
	\includegraphics[width=\linewidth]{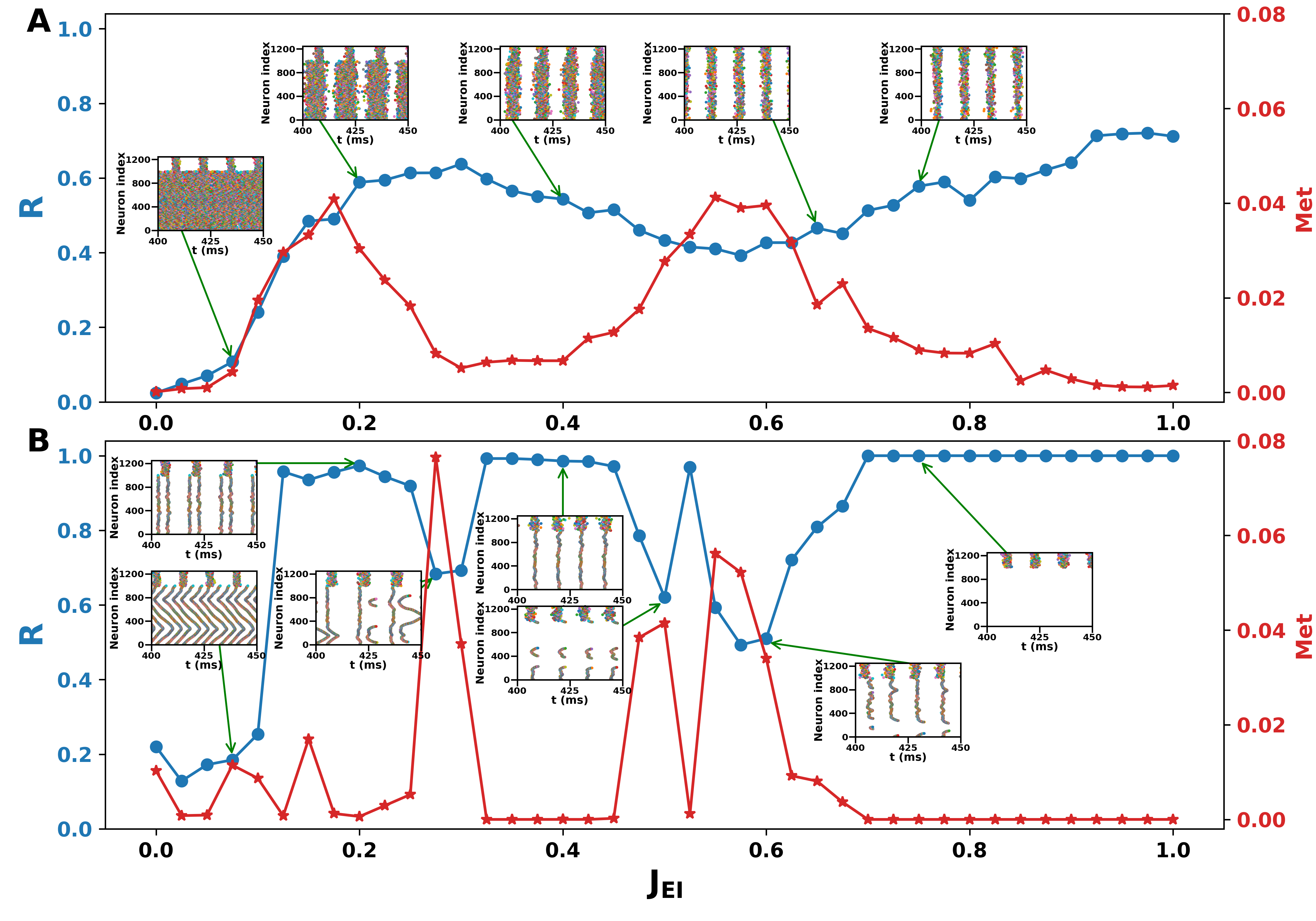}	
	\caption{ Influence of the inhibitory level $J_{EI}$ on the emergence of spatiotemporal dynamical states shown in  \textbf{A}  ( $J_{gap} = 0$)  and \textbf{B}  ( $J_{gap} = 0.1$)  where $J_{EE} = 0.5$.  The emergence of collective dynamics in the excitatory population quantified  by the order parameters ({\it blue circle lines}) and metastability ({\it red star lines}). Inset plots show a detail  of the typical spatiotemporal firing patterns in each dynamics regimes of sweeping parameter  $J_{EI}$.  }
	
	\label{fig:Figure4_FixedJe}
\end{figure}

To clearly distinguish these different types of population behaviors and characterize the effect of inhibition on their
firing patterns,   we compute both order parameter $R$ and metastability $Met$ as a function of inhibitory level $J_{EI}$  in the networks of excitatory population, which  built with (without) gap junction when $J_{EE} =0.5$. The obtained results are presented in \hyperref[fig:Figure4_FixedJe]{Fig.\ref{fig:Figure4_FixedJe}} where each subplot illustrates measures of synchronization and metastability as well as examples of several firing patterns  in each dynamics regime, indicated in the inset plots.  As seen in the \hyperref[fig:Figure4_FixedJe]{Fig.\ref{fig:Figure4_FixedJe}A}, in the network of excitatory population in absence of gap junction, variation of $J_{EI}$  only induces incoherent state, metastable state (it is still irregular but with slight clustering firing pattern) and generalized synchronous state  when  both $R$ and $Met$ satisfy the following conditions: $R\approx 0,Met \approx 0$;  $ 0.4<R<0.8,0 < Met$; $ 0.7<R<0.8,Met \approx 0$, respectively.  It is confirmed again that increasing $J_{EI}$ has a tendency to induce the lower level of synchronization, namely generalized synchronous state.  However, it is seen that the excitatory population built with gap junction, in \hyperref[fig:Figure4_FixedJe]{Fig.\ref{fig:Figure4_FixedJe}B}, exhibits characteristically distinct types of dynamical behaviors as $J_{EI}$ varies.   Note that their
respective dynamics were previously illustrated in  \hyperref[fig:ParaSwape_JeeVsJei]{Fig.\ref{fig:ParaSwape_JeeVsJei}} and \hyperref[fig:Figure3_rasterplot]{Fig.\ref{fig:Figure3_rasterplot}}. More precisely, we observe traveling wave patterns with lower level of inhibition for $J_{EI} \leqslant 0.1$ when $R\approx 0.2, Met < 0.01$.  Excitatory population starts to exhibit synchrony patterns with period 2 as shown inset plot with a slight increase in $J_{EI}$  which is in range of $0.1 < J_{EI}\leqslant 0.25$ when  $R> 0.91, Met < 0.01$. Then, a further increase in inhibitory level to the range of $0.25 < J_{EI}\leqslant 0.3$ when $R$ (around 0.65) is middle-value and $Met$ (greater than 0.07) is high-value,  we observe the metastable state in which the dynamics in excitatory population switches between synchronous and asynchronous state (traveling wave patterns or subthreshold oscillations state). If the inhibitory level is chosen in the range of  $0.3 < J_{EI}\leqslant 0.45$ when $R\approx 1, Met \approx 0$, we find that it gives rise to another synchrony neural activity with period 1.   Subsequently, when $J_{EI}$ lies in the range of $0.45 < J_{EI}\leqslant 0.675$ with middle-value of both $R$ and $Met$,  we see a peculiar chimeraness behavior in which the excitatory population splits into three domains: coherent state, various ripples events (such as traveling waves) and  oscillation without spikes.  Coexistence of these three dynamics states depending on the inhibition level is the first time to be observed.  Eventually, the excitatory populations exhibits subthreshold synchronous states -- oscillation without spikes-- due to stronger inhibitory level in the range of  $0.675 < J_{EI}$.  It is evident that inhibitory level $J_{EI}$ is a significant system parameter which plays a major role in determining the emergence of the above mentioned firing patterns, including traveling wave, synchrony state with period 2 (or 1),  a peculiar chimera-likeness behavior and subthreshold synchronous states. More importantly, there are transitions between dynamical states depending on inhibitory level among individual neurons.

\begin{figure}[H] 	
	\includegraphics[width=\linewidth]{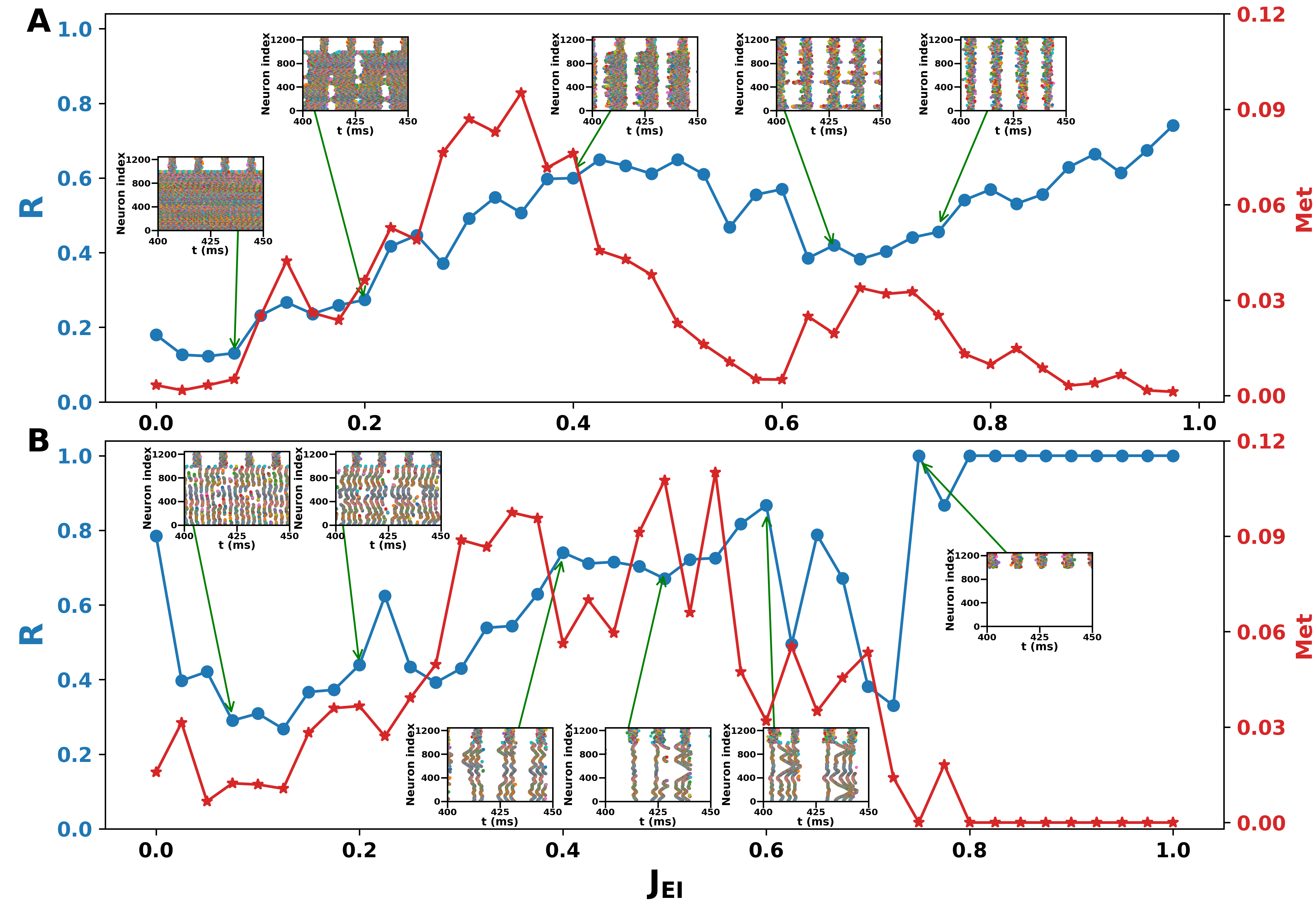}	
	\caption{   Distinct types of dynamical behavior as $J_{EI}$ varies as illustrated  in  \textbf{A} ($J_{gap} = 0$ ) and \textbf{B}  ($J_{gap} = 0.1$) but with high excitatory level where $J_{EE} = 2.5$. Inset plots present typical firing patterns in each dynamics regimes.   }
	
	\label{fig:Figure5_FixedJe25}
\end{figure}

We have so far focused on changes in the inhibitory level to study the emergence of several firing patterns with  fixed weaker  excitability level. In the following, we continue investigating the population dynamics as a functions of inhibitory level $J_{EI}$  but with stronger excitability level. As an  extension of the findings about the effect of inhibitory level in the emergence of various firing patterns, we still consider excitatory population built with (or without) gap junction that are shown in \hyperref[fig:Figure5_FixedJe25]{Fig.\ref{fig:Figure5_FixedJe25}}. For each case, we calculate the order parameter and metastability of  various firing  patterns by varying the inhibitory level $J_{EI}$ for a fixed value of stronger  excitatory strength $J_{EE} = 2.5$. The typical firing patterns of  excitatory population for each dynamics regimes are shown in inset plots of \hyperref[fig:Figure5_FixedJe25]{Fig.\ref{fig:Figure5_FixedJe25}}. As seen in the \hyperref[fig:Figure5_FixedJe25]{Fig.\ref{fig:Figure5_FixedJe25}A}, variation of inhibitory level of the excitatory part  built without gap junction causes the similar modes of firing pattern, which has been shown in \hyperref[fig:Figure4_FixedJe]{Fig.\ref{fig:Figure4_FixedJe}A}. In network built with hybrid synapses, it is seen that there are three  different  types of neuronal activity in the excitatory population obtained as shown in \hyperref[fig:Figure5_FixedJe25]{Fig.\ref{fig:Figure5_FixedJe25}B}. There exists  a much more regular  population activity, however it is still an incoherent state,   for weak inhibitory synaptic connections.  It is because the range of $J_{EI}$ becomes  narrower for  traveling waves and wider for chimera-likeness state  with increasing  excitability levels that are shown in \hyperref[fig:ParaSwape_JeeVsJei]{Fig.\ref{fig:ParaSwape_JeeVsJei}B}. This indicates that traveling waves vanish at very high excitability levels (see \hyperref[fig:ParaSwape_JeeVsJei]{Fig.\ref{fig:ParaSwape_JeeVsJei}B-D} and \hyperref[fig:Figure5_FixedJe25]{Fig.\ref{fig:Figure5_FixedJe25}B}) , which is similar with previous study\cite{calim2018chimera}.   If the inhibitory coupling strength is sufficiently increased  among range $ 0.25<J_{EI}\leqslant 0.7$ where both $R$ and $Met$ are high-value, we observe chimera-likeness behavior in which the excitatory population coexists the synchrony state and traveling waves at the same time. In addition, we also see metastable state  that alternates between synchronous and incoherent behaviors with different time windows. To check the persistence of these two special behavior, we further inscreased $J_{EI}$ and observed that chimera-likeness and metastable state does no longer exist, instead a subthreshold  oscillations (without spikes) emerges for $0.7<J_{EI}$. To sum up,  we show that balanced and hybrid synaptic connected network of Wang-Buzsáki neurons can displays  various types of different levels of synchronous activity that imply vastly different computational properties. For weak excitatory coupling, the network of excitatory population displays rich collective dynamics -- traveling waves, synchrony patterns with period $2$, metastable state, synchrony patterns with period $1$,  chimera-likeness/metastable state and  lastly oscillation without spikes -- as inhibitory level $J_{EI}$ varies. However, for strong excitatory couplings,  we find that the increasing inhibitory level $J_{EI}$ in the networks of excitatory population only leads to three fundamentally different types of neuronal activity, namely, less regular population activity, chimera-likeness behavior/metastable state and subthreshold synchronous states (oscillation without spikes).

\begin{figure}[H] 	
	\includegraphics[width=\linewidth]{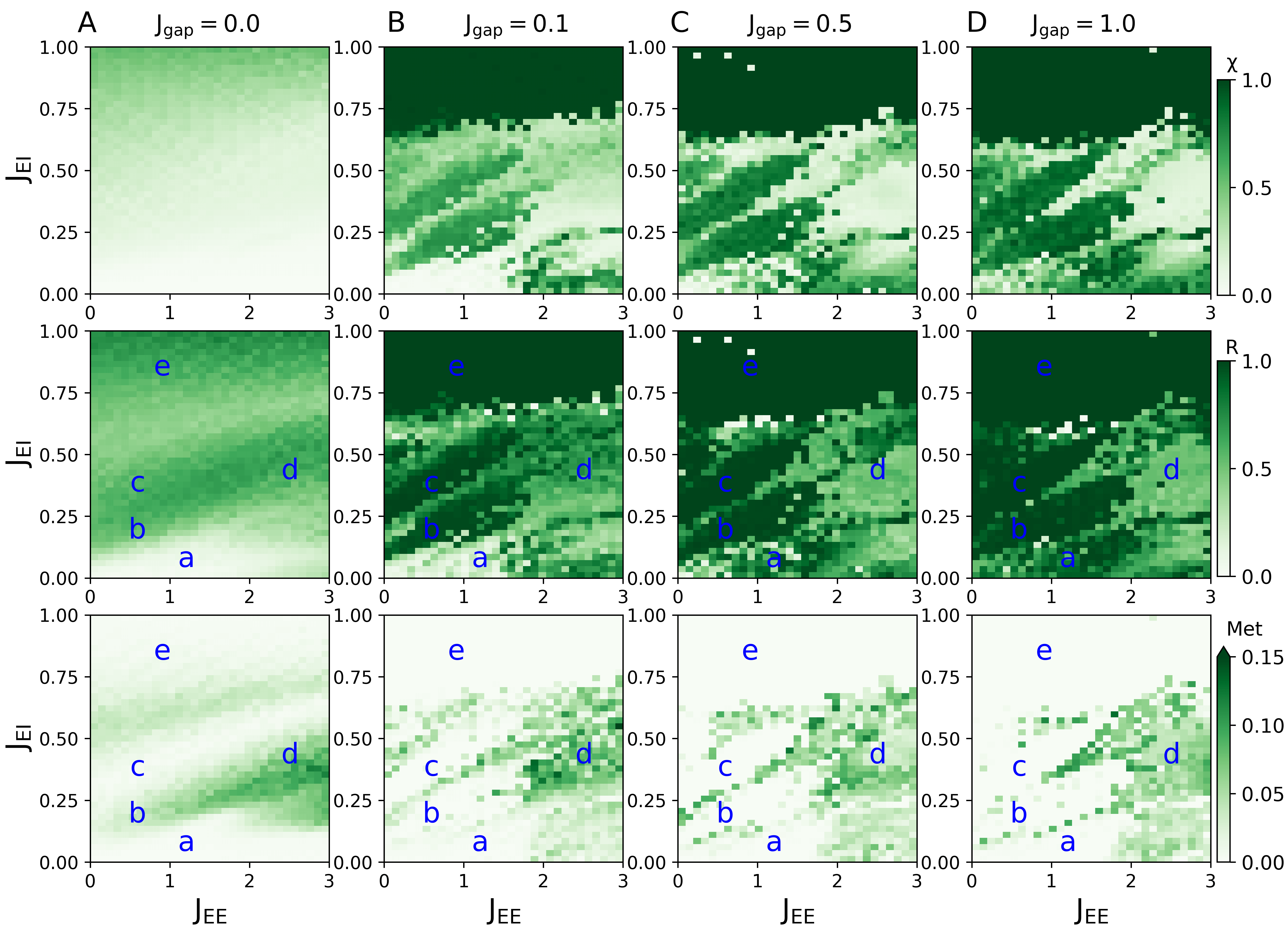}	
	\caption{  Influence of  probability  $P_{sw}$ for small-world topology  on different levels of synchronization and its robustness on  ($J_{EE}, J_{EI}$) plane,  unfolding a clear comparison with \hyperref[fig:ParaSwape_JeeVsJei]{Fig.\ref{fig:ParaSwape_JeeVsJei}}. }
	
	\label{fig:Figure6_ParaSwape}
\end{figure}

\subsection*{\textbf{Influence of  probability  $P_{sw}$  on excitatory population behavior}}
To present a broader perspective on different levels of synchronization and to confirm its robustness,  we choose a  bigger  probability  $P_{sw}$ ($=0.15$) for small-world topology of excitatory synapse connections and show excitatory population behavior on  ($J_{EE}, J_{EI}$) plane as depicted in \hyperref[fig:Figure6_ParaSwape]{Fig.\ref{fig:Figure6_ParaSwape}}. It is clear, in the network built with hybrid synapses, that the bigger probability for small-world topology still supports rich excitatory population behavior emerging as five different dynamical regimes marked by latters $a, b, c, d,e$ -- that is  traveling waves, two types of synchronous states, chimera-likeness or metastable state with various ripples events, and subthreshold synchronous states.  Our results reveal that the role of gap junction and excitatory-inhibitory balance to network states are robust   with network densities. 

\section{Discussion}

We briefly discuss the different characteristics and functionality of these two types of chemical and electrical synapses (For reviews, see\cite{pereda2014electrical,alcami2019beyond}). The most common mechanism for transmission between neurons is chemical synapses, which is  typically unidirectional and relies on the intermittent presence of coming spikes  from presynaptic terminal to generate transient changes. From this point, chemical synaptic communication is episodic. In addition, neurotransmitter release is probabilistic, and failures in transmission will occasionally occur despite the presence of coming spikes from presynaptic terminal. Depending on the neurotransmitter, a chemical synapse can either be excitatory or inhibitory. By contrast,  electrical synaptic transmission through gap junctions underlies direct, continuous, reliable and rapid neuronal communication in the CNS. Electrical synapses are bidirectional and of local connections between neurons which are spatially very close.  The gap junctional conductance is remarkably and characteristically constant for sustained periods of time.  However, recent studies had evidenced that both chemical and electrical modalities of synaptic transmission are plastic and capable of modifying their strength in a bidirectional fasion. For example, activity-dependent changes in the strength of chemical synapses are thought to underlie mechanisms of learning and memory (for a review, see ref \cite{Martin2000}).  Regarding to the functional role of gap junction plasticity, such as changes in light conditions\cite{bloomfield2009diverse},  has been experimentally demonstrated \cite{Robinson2017,Wang7616,TURECEK20141375,COULON20171275,pernelle2018gap} (also reviewed in refs \cite{pereda2014electrical,Haas2016}). 

The distribution and role of electrical synapses  have been investigated in the previous studies\cite{Galarreta2001,Connors2004}. A series of reports had demonstrated that electrical synapses occur between excitatory projection neurons of the inferior olivary nucleus and  between interneurons in several regions of the adult brain, including the cerebellum, neocortex, hippocampus and so on (see review\cite{Galarreta2001}  detailly). Computer simulations had greatly contributed to revealing the impact and  functional role of gap junctions between inhibitory interneurons on network activity. Firstly,  gap junctions between inhibitory interneurons have been shown to enhance synchronized neuronal firing\cite{Connors2004,Pernelle2018}, which is one of the most distinct and recognizable properties of the electrical synapses.  Secondly, it has been suggested that the presence of electrical synapses could mediate close synchronization of subthreshold and spiking activity among clusters of neurons\cite{alcami2019beyond,Placantonakis5008}. In all, the role of electrical synapses for network activity has been the topic of various review articles (see Ref\cite{Galarreta2001,Connors2017}) and therefore will not be extensively discussed here. The influence of electical coupling for balanced networks on generation of distinct dynamical regimes has been investigated in this paper. Introducing the gap junction, even with weaker coupling weights,  to excitatory population leads to  various firing patterns -- such as traveling wave, synchronous states etc.  However, gap junctions are often  observed between inhibitory neurons. The role  and function of gap junctions between inhibitory neurons,  in the  generation of spatiotemporal firing patterns,  will be the subject for the future work.

Coactivated inhibitory and excitatory synaptic inputs plays a role in diversity of cortical functions and general information processing\cite{Buzsaki2006network,RubinE93662017}.  It is generally accepted that activity of excitatory networks, if only excitatory cells were present in networks, is simple and predictable -- excitation just leads to futher excitation. Inhibitory networks are totally different\cite{ISAACSON2011231}, namely, it can generate nonlinear effects. In the absence of inhibition, any type of external input,  would generate more or less the same one-way patterns. Theoretical work has shown that networks built from both excitatory and inhibitory elements can self-oranize and generate complex properties\cite{anderson2000orientation,ATALLAH2009566,Wehr2003,Cindy2009Odor}.  For example, study of Rubin et al.\cite{RubinE93662017} shown that the ratio of excitation-inhibition balance is essential for robust neuronal selectivity and crucial for stability in associative memory networks in the presence of noise.

In this work,  we investigate the emergence of various firing patterns in excitation and inhibition balance  networks of itself excitatory cells coupled via hybrid synapses. Our findings  in this  class of network illustrate how level of inhibition can synchronize or desynchronize, detect coincidences, inhibit on population behavior. More detail,  with weaker excitability level,  variation of inhibitory level for excitatory population leads to  rich collective dynamics, such as traveling waves, synchrony and desynchrony state (chimera-likeness or metastable state) and oscillation without spikes due to stronger inhibition.   With stronger excitability level, increasing inhibitory level for the same networks causes three fundamentally different types of neuronal activity --  less order population activity, chimera-likeness behavior (or metastable state  at some times) and subthreshold synchronous states.

Brain dynamics is often inherently variable and unstable, consisting of sequences of transient spatiotemporal patterns\cite{sporns2010networks,Rabinovich2008}. These squences of transients are a hallmark of metastable dynamics that are neither entrely stable nor completely unstable. Our results pay a possible way  to uncover the underlying mechanisms of generating metastable dynamics, such as chimera-likeness, that may mediate perception and cognition.

\begin{acknowledgements}
We thank the funding from Fondecyt Project Nos.3170342 (K.X.) and 1181076 (P.O.) from CONICYT, Chile. P.O. is partially funded by the Advanced Center for Electrical and Electronic Engineering (FB0008 Conicyt, Chile). The Centro Interdisciplinario de Neurociencia de Valparaíso (CINV) is a Millennium Institute supported by the Millennium Scientific Initiative of the Ministerio de Economía (ICM-MINECON, Proyecto P09-022-F,CINV, Chile).
\end{acknowledgements}

%
 \section*{Conflict of interest}
The authors declare that they have no conflict of interest.

\bibliographystyle{unsrt}

\bibliography{DiveMixedSyn}   

%
%

\end{document}